\documentclass[showpacs,aps,prl,twocolumn,floatfix]{revtex4}

\usepackage{times}
\usepackage{graphicx}
\usepackage{bm}
\usepackage{amsmath}

\newcommand{\s}{\sum\limits}

\newcommand{\pa}{\partial}

\newcommand{\be}{\begin{equation}}
\newcommand{\e}{\end{equation}}
\newcommand{\beml}{\begin{subequations}}
\newcommand{\eml}{\end{subequations}}
\newcommand{\beq}{\begin{eqnarray}}
\newcommand{\eq}{\end{eqnarray}} 
\newcommand{\ba}{\begin{array}}
\newcommand{\ea}{\end{array}}
\newcommand{\lt}{\left}
\newcommand{\rt}{\right}
\newcommand{\n}{\nonumber}
\newcommand{\la}{\langle}
\newcommand{\ra}{\rangle}

\newcommand{\im}{\,{\rm Im}\,}
 
\newcommand{\ep}{\varepsilon} 
 
\DeclareMathOperator{\var}{Var}

\begin{document}
 
\date{31 March, 2005}

\title{Non-universality of Anderson localization 
in short-range correlated disorder}

\author{M. Titov$^{(1)}$ and H. Schomerus$^{(1,2)}$}

\affiliation{{}$^{(1)}$ Max-Planck-Institut f\"ur Physik komplexer Systeme,
N\"othnitzer Str. 38, 01187 Dresden, Germany
\\
{}$^{(2)}$
Physics Department,
Lancaster University,
Lancaster, LA1 4YB, UK
}

\begin{abstract}

We provide an analytic theory of Anderson localization on a lattice
with a weak short-range correlated disordered potential.
Contrary to the general belief we demonstrate 
that already next-neighbor statistical correlations in the potential
can give rise to strong anomalies in the localization length and 
the density of states,
and to the complete violation of single parameter scaling.
Such anomalies originate in additional symmetries of the lattice model in
the limit of weak disorder. The results of numerical simulations 
are in full agreement with our theory,
with no adjustable parameters.

\end{abstract}

\maketitle

It is customary to assume that a wave completely looses 
its phase memory when reflected from a weak 
disordered potential \cite{Ishimaru,Sheng,Berkovits,Genack}.
Many powerful analytical frameworks, 
such as the single-parameter scaling theory  \cite{ATAF},
the DMPK approach \cite{D,MPK}, 
or 1D non-linear $\sigma$-model \cite{Efetov} are based, 
implicitly or explicitly, on the reflection phase randomization (RPR). 
The RPR stands behind the notion of the standard universality classes 
in the random matrix description of non-interacting disordered wires
\cite{CWJreview} and leads to the independence of the mean density of states 
on disorder strength.

The violation of the RPR 
is responsible for many non-universal effects. 
One example is the tight-binding model with hopping disorder, where
the density of states diverges at the band center
for an odd number of coupled chains \cite{Dyson}, 
while it vanishes for an even number \cite{BMF2000}.
The breaking of single parameter scaling and RPR by hopping disorder
and other deviations from universality were studied recently 
in great detail \cite{RMF2004,ST2003b}.

A partial violation of the RPR can be induced by a disordered 
on-site potential alone.
This happens when the potential can no longer be regarded as weak, 
e.g., at the edges of electronic conduction (or photonic 
transmission) bands \cite{ST2003a,DEL},
or in the situation when the probability density of the potential 
has power law tails \cite{ST2003c}.
In the case of a one-dimensional lattice with 
white-noise potential the RPR is partially violated at the band center,
leading to the Kappus-Wegner anomaly characterized by a
9 \% increase of the localization length \cite{KW,Derrida}.

In this paper we demonstrate that the RPR can be broken 
far more dramatically if the disordered potential 
is short-range correlated. (By the short range we mean
a finite correlation radius which is much smaller than the localization length.)
The lack of RPR is accompanied by anomalies in the 
localization length (which can sharply increase or decrease),
and in other quantities such as the delay time or the density of states. 
In contrast to the general belief \cite{OP1997,DEL2003}
even next-neighbor statistical correlations in the potential 
can lead to a severe violation of the single parameter scaling.
In brief we distinguish two different effects of the correlations:
(i) the system retains its universal properties with a
renormalized localization length \cite{IK};
(ii) the universality is broken in the vicinity of specific spectral points;
the RPR and the single parameter scaling are violated; 
the density of states shows an anomaly, which depends on disorder strength.

\begin{figure}[t]
\includegraphics[width=0.9\columnwidth]{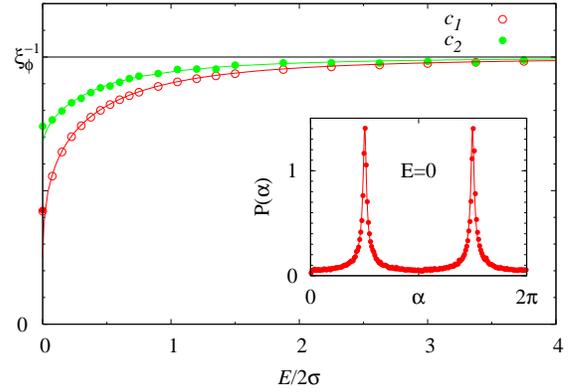}
\caption{A lack of phase randomization 
leads to a distinctive band center anomaly in
the conductance $g_n$ through a disordered wire (\ref{A}) of length $n=12000$
with next-neighbor
correlations  (\ref{BCanomaly}) of the potential
and $\sigma=1/150$ (RPR localization length $\xi_\phi=2/\sigma=300$).
The main panel shows the mean logarithm
$c_1=-(1/2n)\la\ln g_n \ra$ and its variance
$c_2=(1/4n)\var \ln g_n$ from numerical simulations
(data points) 
and the analytical expressions 
(\ref{c1},\ref{c2}) (solid lines), scaled to
$\xi_\phi^{-1}$. The inset shows the distribution of 
the reflection phase $P(\alpha)$ at the band center,
calculated numerically 
as well as analytically (\ref{BCdistribution}).
}
\label{fig:BCanomaly}
\end{figure}

We consider the one-dimensional Anderson model 
\be
\label{A}
-\Psi_{n+1}-\Psi_{n-1}+U_n \Psi_n = E\,\Psi_n
\e
with a weak disordered potential $U_n$,
$\la U_n \ra =0$ with $|U_n|\ll 1$. 

The correlation function $\la U_n U_m\ra =\sigma(n,m)$ is assumed 
to be invariant under a finite translational shift 
$\sigma(n+Q,m+Q)=\sigma(n,m)$. The minimal period 
$Q$ does not need to be identical with the lattice constant.
This accommodates the cases of structural and chemical disorder, 
or carefully engineered disorder, e.g., with masks. 
For $Q=1$ the correlation function is generally written as
\be
\label{InvariantDisorder}
\la U_n U_{n'} \ra =\sigma_{n-n'}, \qquad \sigma_n=\sigma_{-n}. 
\e

The localization length $\xi$ is accessible via the 
dimensionless conductance (transmission probability) $g_n$ of the
system of length $n$, $\xi^{-1}=-\lim_{n\to\infty}(1/2n)\la\ln g_n\ra$.
In order to quantify deviations from universality
we consider the complete statistics of the conductance
fluctuations obtained from the generating function
\be
\label{GenFunc}
\chi(\mu)=\lim_{n\to\infty}\frac{1}{n} \ln \la g_n^{-\mu/2}\ra
=\s_{s=1}^\infty \frac{c_s \mu^s}{s!}, \quad \mu\geq 0.
\e
The localization length is given by $\xi=c_1^{-1}$. In
single-parameter scaling \cite{ATAF},
$c_2=c_1$, and $c_s=0$ for $s\geq 3$, 
corresponding to a log-normal distribution of $g_n$.
In this paper we concentrate on the first two coefficients
$c_1$ and $c_2$. We base our analysis on the exact phase 
formalism \cite{Halperin,LGP}, 
which we extend to the case of correlated disorder.

For the sake of definiteness assume the wave is reflected 
from the right end of a system of length $n\gg\xi$, 
with reflection amplitude $r_n=\sqrt{1-g_n}\exp(i\alpha_n)$.
The reflection phase $\alpha_n\in(0,2\pi)$ is related to the wave function by 
\be
\label{RicVariable}
\frac{\Psi_n}{\Psi_{n-1}}=\frac{1+e^{i\alpha_n}}{e^{i k}+e^{-i k+i\alpha_n}},
\quad E=-2\cos k,
\e
while the conductance $g_n$ is obtained from
\be
\label{Lambda}
\lambda_{n}\equiv -\frac{1}{2}\ln g_n
=\frac{1}{2}\ln|\Psi_{n}^2+\Psi_{n-1}^2+E\,\Psi_{n}\Psi_{n-1}|.
\e
Up to the second order in disorder strength,
the statistical evolution of the phase and conductance
with increasing system size is described by 
the recursion relations
\beml
\beq
\label{ReccPhase}
\alpha_{n+1}-\alpha_n &=& 2k-2 K_n(\alpha_n),\\
\label{ReccLambda}
\lambda_{n+1}-\lambda_n &=&
K'_n(\alpha_n)+
\big(K'_n(\alpha_n) \big)^2,
\eq
\eml
where the prime stands for the derivative with respect to $\alpha_n$ and
the function $K_n(\alpha)$ is given by
\be
\label{K}
K_n(\alpha)=\frac{\Phi_n(\alpha)}{1+
\Phi'_n(\alpha)},\quad
\Phi_n(\alpha)=\frac{U_n}{v}(1+\cos \alpha),
\e
with the group velocity $v=|\pa E/\pa k|$.

There exsist two major obstacles in the derivation of
the Fokker-Planck equation for the 
joint probability density $P_n(\alpha,\lambda)$ from
Eqs.~(\ref{ReccPhase},\ref{ReccLambda}).
First, the wave-number $k$ on the right-hand side
of Eq.~(\ref{ReccPhase}) is not small. 
Second, the values of $K_n$ are correlated at different sites.
These difficulties can be circumvented by monitoring
the variables $\alpha_n$ and $\lambda_n$ with a step of
$q$ sites \cite{Lambert}. We parameterize the energy  
$E=-2\cos \pi p/q+\ep$
with small $\ep\ll 1$ and integer $p$, and
choose $q$ to be much larger 
than the correlation radius of the potential. 
Thus, the change of the phase over $q$ sites 
$\delta\alpha=\alpha_{n+q}-\alpha_n$ is small,
\beq
\label{dPhase}
&&\delta\alpha_n= \frac{2\ep q}{v}
-2\s_{s=0}^{q-1}\Phi_{n+s}(\alpha+2 ks)\\ \n
&&\quad + 2\s_{s=0}^{q-1}\s_{m=s}^{q-1}\frac{2}{1+\delta_{ms}}
\Phi_{n+s}(\alpha+2ks)\Phi'_{n+m}(\alpha+2km),
\eq
to the second order in the potential.
Similarly, the increment $\delta\lambda=\lambda_{n+q}-\lambda_n$
is given by
\beq
\label{dLambda}
&&\delta\lambda_n=\s_{s=0}^{q-1}\Phi'_{n+s}(\alpha+2 ks)\\ \n
&&\quad - \s_{s=0}^{q-1}\s_{m=s}^{q-1}\frac{2}{1+\delta_{ms}}
\Phi_{n+s}(\alpha+2ks)\Phi''_{n+m}(\alpha+2km).
\eq
In the limit o weak disorder 
the recurrent relations Eqs.~(\ref{dLambda},\ref{dPhase}) 
lead to the Fokker-Planck equation for
the joint probability density $P_n(\alpha,\lambda)$
\beq
\label{FPgeneral}
&&\frac{\pa P_n}{\pa n}=-\frac{\pa}{\pa \alpha}F(\alpha)P_n
+\frac{1}{2}\frac{\pa^2}{\pa\alpha^2}D(\alpha)P_n\\
&&-\frac{\pa}{\pa \lambda}F_1(\alpha)P_n
+\frac{1}{2}\frac{\pa^2}{\pa\lambda^2}D_1(\alpha)P_n
+\frac{\pa}{\pa\lambda}\frac{\pa}{\pa\alpha}D_{01}(\alpha)P_n.\n
\eq
The drift and diffusion coefficients, which determine the
phase distribution $P_n(\alpha)$, are 
\be
\label{Def}
F(\alpha)=\frac{1}{q}\la \delta\alpha_n \ra, \ \
D(\alpha)=\frac{1}{q}\s_{m=-\infty}^\infty\!\! 
\la \delta\alpha_n \delta\alpha_{n+mq}\ra.
\e
In the case of the correlated disoder 
(\ref{InvariantDisorder})
with $Q=1$, 
these coefficients are related by
$F=2\ep/v+(1/4)\pa D/\pa \alpha$. 
The other coefficients in Eq.\
(\ref{FPgeneral})  are given by
\beq
\label{Def1}
&&F_1(\alpha)=\frac{1}{q}\la \delta\lambda_n \ra, \ \
D_1(\alpha)=\frac{1}{q}\s_{m=-\infty}^\infty\!\! 
\la \delta\lambda_n \delta\lambda_{n+mq}\ra.\n\\
&& D_{01}(\alpha)=\frac{1}{q}\s_{m=-\infty}^\infty\!\! 
\la \delta\alpha_n \delta\lambda_{n+mq}\ra.
\eq
Thus, we have derived Fokker-Planck equations 
(\ref{FPgeneral}),
which describe the system in the vicinity of a given rational energy
$E=-2\cos \pi p/q$ with $q\geq 2$. 
The potential fluctuations are assumed to be restricted within the
conduction band, so that $|E\pm U_n|<2$.
In particular Eq.~(\ref{FPgeneral}) does not apply in the 
vicinity of the band edge, where any fluctuation 
is strong. The latter case has to be treated separately.
The effect of dichotomic correlated
disorder near the band edge was studied 
in Ref.~\cite{DEL2003b}.

The equation for the phase distribution function $P_n(\alpha)$
is readily obtained by integrating Eq.~(\ref{FPgeneral}) over 
the variable $\lambda$. There exists, however, no general way
to derive a similar equation for the distribution $P_n(\lambda)$.
The calculation of the generating function $\chi(\mu)$ hence
requires the analysis of the full density $P_n(\alpha,\lambda)$. 
Such analysis is greatly simplified for RPR, which 
implies the factorization $P_n(\alpha,\lambda)=(2\pi)^{-1}P_n(\lambda)$.
In this case a closed equation
for $P_n(\lambda)$ can be derived by averaging 
Eq.~(\ref{FPgeneral}) over the phase 
\beml
\beq 
\label{FPrandomPhase}
\frac{\pa P_n(\lambda)}{\pa n}&=&\frac{1}{\xi_\phi}
\frac{\pa P_n(\lambda)}{\pa \lambda}+\frac{1}{2\xi_\phi}\frac{\pa^2 P_n(\lambda)}{\pa\lambda^2},\\
\label{xiPhi}
\frac{1}{\xi_\phi}&=&\int_{0}^{2\pi}\frac{d\alpha}{2\pi}F_1(\alpha)=
\int_{0}^{2\pi}\frac{d\alpha}{2\pi}D_1(\alpha),
\eq
\eml
where $\xi_\phi$ is the localization length in the presence of RPR.
The last equality in Eq.~(\ref{xiPhi})
follows directly from Eq.~(\ref{ReccLambda}), which implies that
the first and second moment of the increment of $\lambda$ are equal
if the phase is randomized. 
The generating function (\ref{GenFunc}) calculated from Eq.~(\ref{xiPhi})
has the parabolic shape $\chi(\mu)=\mu(1+\mu/2)/\xi_\phi$.
Hence, RPR implies single-parameter scaling in the localized regime,
even in the presence of correlated disorder [this statement holds also
for hopping disorder, which modifies the expression for $K_n$, 
while Eqs.~(\ref{ReccPhase},\ref{ReccLambda}) retain their form.]

It is instructive to calculate $\xi_\phi$ 
for correlations of the type (\ref{InvariantDisorder}). 
Taking the integrals in Eq.~(\ref{xiPhi}) we reproduce 
the result of Ref.~\cite{IK} 
\be
\label{PerturbativeMFP}
\frac{1}{\xi_\phi}=
\frac{1}{2v^2}\s_{s=-\infty}^{\infty}\sigma_s e^{2i k s}.
\e
It is important to remember, however, that $\xi_\phi=\xi$ only if RPR
holds. This brings us to the question:
What are the conditions for RPR, and  what are the implications
for the localized wave functions when RPR does not occur?

As a first example to illustrate the effect of short-range correlations
on the reflection phase statistics, 
we consider disorder of the type (\ref{InvariantDisorder}) with
next-neighbor correlations only, $\sigma_{s\geq 3}=0$. 
We let $q=2p, q\gg 1$ and obtain from Eq.~(\ref{Def})
\be
D(\alpha)=2\sigma_0-(\sigma_0-2\sigma_1)\sin^2\alpha.
\e
The Fokker-Planck equation for the stationary phase 
distribution $P(\alpha)$ is simplified to
\be
\label{FPalpha}
-\ep\frac{\pa}{\pa \alpha}P
+\frac{1}{2}\frac{\pa}{\pa\alpha}\sqrt{D}\frac{\pa}{\pa\alpha}\sqrt{D}P=0.
\e
Its solution at the band center $\ep=E=0$ is given by
$P\propto [D(\alpha)]^{-1/2}$. 
Note, that the solution slightly deviates from the constant even 
for the white-noise potential ($\sigma_1=0$),
which is the source of the so-called 
Kappus-Wegner anomaly \cite{KW,Derrida,ST2003a}.
The next-neighbor statistical correlations can induce 
far stronger deviations from RPR. Indeed, 
for $\sigma=\sigma_1=-\sigma_0/2$,
\be
\label{BCanomaly}
\la U_n U_{n'}\ra =2\sigma\delta_{nn'}-\sigma\delta_{nn'\pm 1},
\e
the solution to Eq.~(\ref{FPalpha}) is singular at $\ep=0$, 
because the function $D(\alpha)$ develops a zero for $\alpha=\pm\pi/2$.

At the level of Eq.~(\ref{FPalpha}) the situation is analogous to the
band center Dyson singularity \cite{Dyson} in the presence of
hopping disorder. Hence, the correlations turned a weak anomaly into a
strong anomaly. While the usual Dyson singularity appears as a consequence of an exact
``chiral'' symmetry of the wire,
such symmetry is limited to the first two orders in disorder strength  
for the correlated disorder (\ref{BCanomaly}).
Using the decomposition $U_n=u_n-u_{n-1}$ with 
$\la u_n u_{n'}\ra=\sigma\delta_{nn'}$, and taking also the fourth order terms
$\la u_n^4\ra=\eta^2$ in the potential into account,
we derive at the band center the regularized phase distribution
\be
\label{BCdistribution}
P_{E=0}(\alpha)\propto \lt(\cos^2\alpha+(\eta^2-\sigma^2)/4\sigma
\rt)^{-1/2},
\e
which becomes more singular for lower disorder strength. 
The probability density $P_{E=0}(\alpha)$ acquires a non-universal 
dependence on the shape of the distribution function of the potential 
via the relation between its fourth and second moment.
Slightly away from the band center, 
$\ep \gg \sqrt{\sigma(\eta^2-\sigma^2)}$, 
the fourth-order terms play no role and
the phase distribution is described by the solution of 
Eq.~(\ref{FPalpha}) with $D=4\sigma \cos^2\alpha$,
\be
\label{solve}
P(\alpha)\propto\int_{0}^\infty\!\!dy\,\frac{e^{-\ep y/2\sigma}}
{\sqrt{y^2\cos^2\alpha+y\sin 2\alpha+1}}.
\e

The lack of RPR is necessarily reflected 
in an anomaly of the density of states, due to
the node-counting theorem \cite{Schmidt,Buettiker1993}; 
this also implies a different
statistics of time delays  $\tau=\pa \alpha/\pa E$, 
for which the increments are obtained
by differentiating Eq.~(\ref{dPhase}) with respect to $\ep$.

\begin{figure}[t]
\includegraphics[width=0.9\columnwidth]{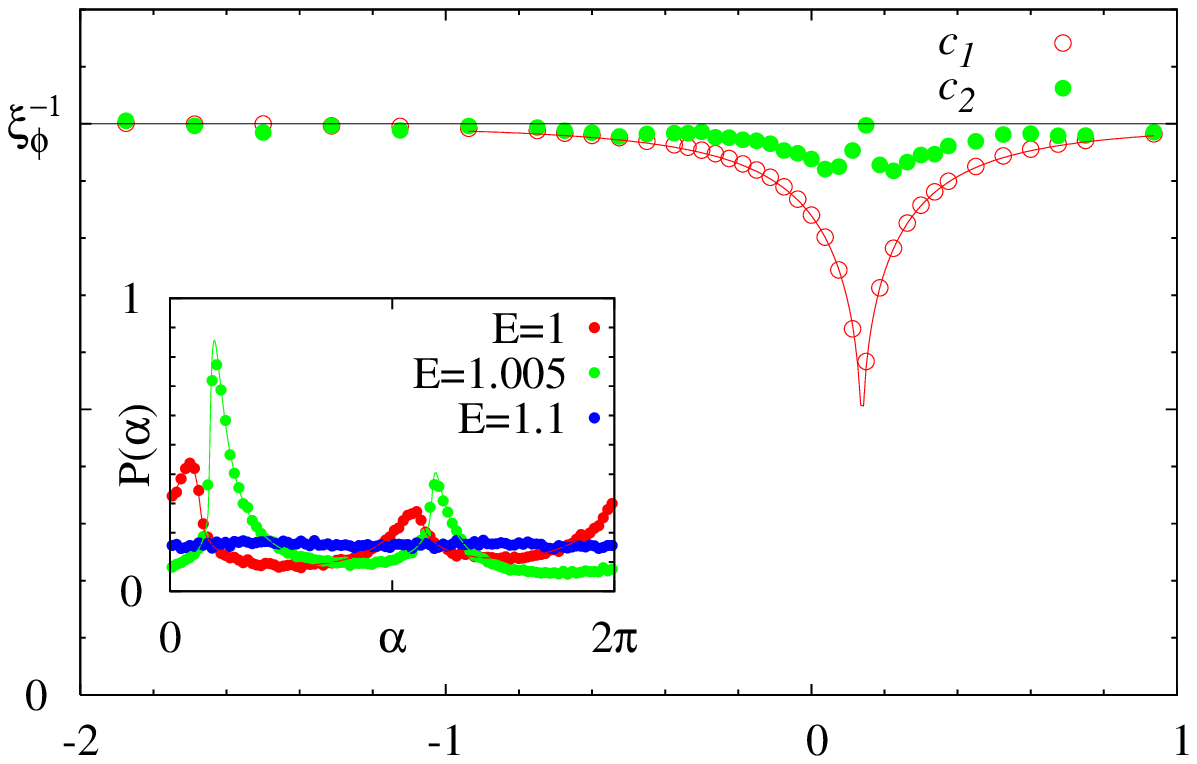}
\includegraphics[width=0.9\columnwidth]{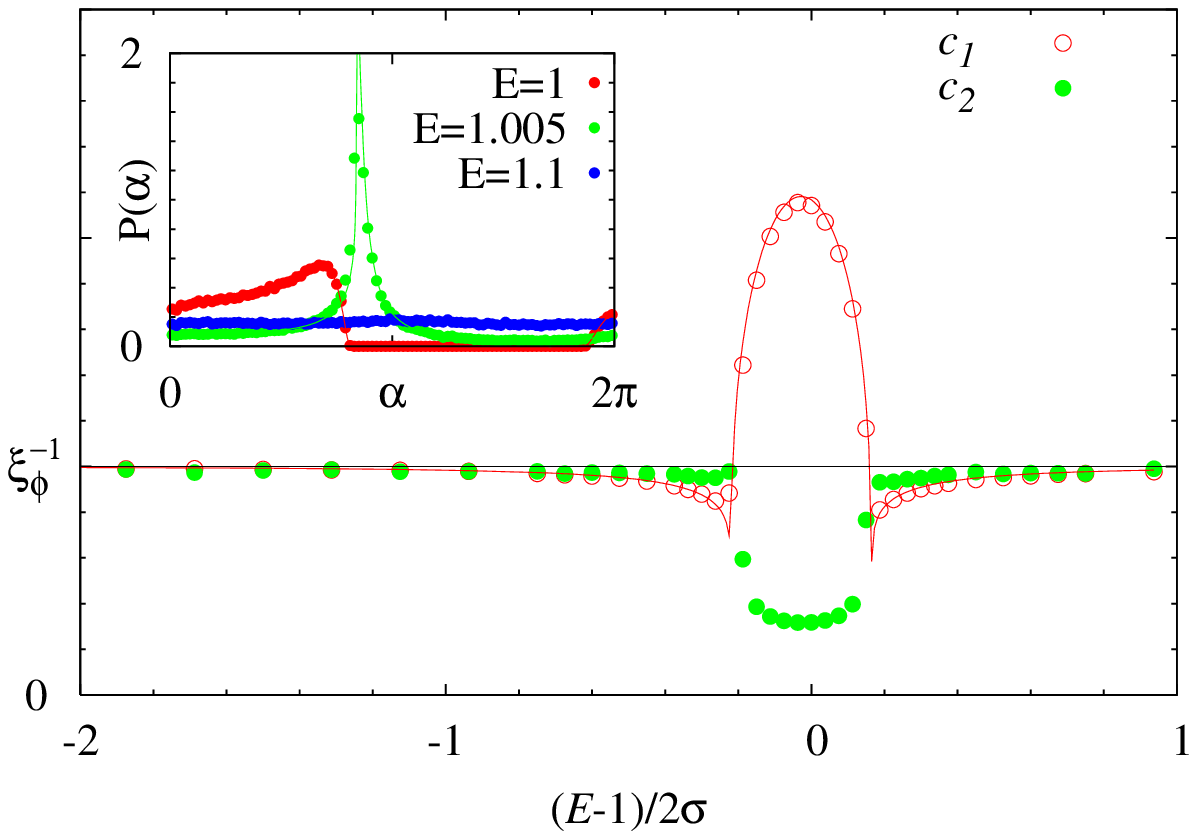}
\caption{The mean logarithm of the conductance 
and its variance are calculated numerically 
for the system (\ref{A}) of the large length $n$.
The coefficients $c_1=-(1/2n)\la\ln g_n\ra$ 
and $c_2=(1/4n)\var \ln g_n$ are plotted as the 
function of energy near the quarter band $E=1$. 
The disorder is generated by $U_n=2u_n-u_{n+1}-u_{n-1}$
(top panel) and $U_n=2u_{n+1}-u_n-u_{n-1}$
(bottom panel), where the random numbers $u_n=0$
unless $n$ is a multiple of $3$ and 
$\la u_{3m}u_{3m'}\ra=\sigma\delta_{mm'}$, 
$\sigma=1/150$.
The insets show the distribution function of the
reflection phase. The solid lines are obtained from 
the solution of Eq.~(\ref{FPgeneral}) with $p=-1, q=3$. 
The RPR localization length $\xi_\phi$ 
is obtained from Eq.~(\ref{xiPhi}).
}
\label{fig:SpecialDisorder}
\end{figure}

We now explore the implications for the transmission properties of
the system. For the specific correlations (\ref{BCanomaly}),
the Fokker-Planck equation (\ref{FPgeneral}) simplifies 
near $E=0$ by
\be
\label{FPal}
\frac{\pa P_n}{\pa n}=
-\ep\frac{\pa P_n}{\pa \alpha}+2\sigma
\lt[\frac{\pa}{\pa\alpha}\cos\alpha+\frac{1}{2}\sin\alpha
\frac{\pa}{\pa\lambda}\rt]^2P_n.
\e
The coefficient $c_1$ is obtained directly by averaging
the expression (\ref{dLambda}) 
with the stationary phase distribution (\ref{solve}). 
The result is 
\be
\label{c1}
c_1=\sigma\int_0^{2\pi}\!\!\!d\alpha\,P(\alpha)\,\cos^2\alpha=
-\frac{\ep}{2} \im \frac{K_1(i\ep/2\sigma)}{K_0(i\ep/2\sigma)},
\e
where $K_\nu(x)$ is the modified Bessel function.
The singularity in Eq.~(\ref{c1}) in the limit $\ep \to 0$
is again regularized in the fourth order of the potential.
In the weak disorder limit we find at the band center
$\xi/\xi_{\phi}=\ln 8\sqrt{\sigma/(\eta^2-\sigma^2)}$. 

From Eq.~(\ref{FPal}) we can determine all coefficients $c_s$ 
recursively \cite{ST2003b,ST2003a,ST2003c}.
A double Laplace transform 
\be
\tilde{P}_\chi(\alpha,\mu)=
\s_n\int_{-\infty}^\infty\!\!d\lambda\, e^{\lambda\mu-\chi(\mu) n}
P_n(\alpha,\lambda),
\e
reduces Eq.~(\ref{FPal}) to the eigenvalue problem
\be
\label{FPeigenvalue}
-\ep\frac{\pa \tilde{P}_\chi}{\pa \alpha}+2\sigma
\lt[\frac{\pa}{\pa\alpha}\cos\alpha
-\frac{\mu}{2}\sin\alpha\rt]^2\tilde{P}_\chi=
\chi(\mu)\tilde{P}_\chi.
\e
The generating function $\chi(\mu)$ can be obtained 
perturbatively in $\mu$, taking
the solution $P(\alpha)$ of Eq.~(\ref{solve}) as zero approximation. 
In particular, the variance $c_2$ is given by
\beml
\beq
\label{c2}
c_2&=&\sigma-c_1-\int_0^{2\pi}\!\!\!d\alpha\, (c_1-\sigma\cos^2\alpha)
G(\alpha),\\
G(\alpha)&=&{\cal L}_\alpha^{-1}\,(c_1\!-\!\sigma\cos^2\alpha\!+\!\sigma
\frac{\pa}{\pa\alpha}\sin2\alpha)P(\alpha),\\
{\cal L}_\alpha&=&-\frac{\ep}{2\sigma}\frac{\pa}{\pa\alpha}+
\frac{\pa}{\pa\alpha}
\cos\alpha\frac{\pa}{\pa\alpha}\cos\alpha.
\eq
\eml

In order to illustrate our predictions, we compare them in
Fig.~\ref{fig:BCanomaly} to the results of numerical simulations.
The numerical results agree with the theory,
without any adjustable parameter.
The main panel shows the energy dependence of $c_1$ and $c_2$ 
near the band center [Eqs.~(\ref{c1},\ref{c2})],
which clearly deviates from the single-parameter
scaling prediction $c_1=c_2=\xi_\phi^{-1}$.
The inset shows the phase distribution 
(\ref{BCdistribution}) at the band center.

In general, RPR is completely violated 
if the diffusion coefficient $D(\alpha)$ has zeros as the function of $\alpha$.
The reflected wave then mantains a strict phase 
relation with the incoming wave.
For the correlated disorder of the type (\ref{InvariantDisorder})
such relation (phase selection) can only occur at the band center.
The restriction is lifted for correlations with $Q\neq 1$.
In Fig.~(\ref{fig:SpecialDisorder}) we provide examples of 
a quarter band anomaly, caused by a disorder correlations with $Q=3$.
Following this recipe, strong anomalies can be produced in a vicinity of 
arbitrary rational values of the wave length $2 q/p$ by a suitably correlated
weak disorder potential with $Q=q$. This is in striking contrast 
to the case of
white-noise disorder, which produces only a weak anomaly, and only
at a single spectral point (the band center).

In conclusion we show that
the 1D Anderson model with a weak short-range correlated potential
may demonstrate strong anomalies near specific spectral points.
Such anomalies can be used to strongly modify the transmission 
properties of electronic wires and photonic wave guides 
in very small energy windows
provided the phase coherence preserved over a long distance.
In these windows the reflected wave develops a preferred phase
relation with the incoming wave. These properties indicate that 
disorder correlations may be favorably employed in the design of
photonic or electronic filters.

We gratefully acknowledge discussions with Piet Brouwer
and the warm hospitality in 
the Lorentz Institute in Leiden, where part of this work was done.

\end{document}